\documentclass[twocolumn,aps,pre,showpacs]{revtex4}
\usepackage[latin1]{inputenc}
\usepackage{graphicx}
\usepackage{latexsym}
\usepackage{amssymb}
\usepackage{amsmath}

\def\d{{\rm d}}
\def\ex{{\rm e}}
\def\y{{\bf y}}
\def\t{{\bf t}}
\def\x{{\bf x}}

\def\z{{\bf z}}
\def\k{{\bf k}}

\def\btau{{\boldsymbol{\tau}}}

\def\beq{\begin{eqnarray}}
\def\eeq{\end{eqnarray}}
\begin{document}

\title{Demographic fluctuations in a population of anomalously diffusing individuals}
\author{Piero Olla}
\affiliation{ISAC-CNR and INFN, Sez. Cagliari, I--09042 Monserrato, Italy.}
\date{\today}

\begin{abstract}
The phenomenon of spatial clustering induced by death and reproduction 
in a population of anomalously diffusing individuals 
is studied analytically. The possibility of social behaviors
affecting the migration strategies has been taken into exam, in the 
case anomalous diffusion is produced by means  
of a continuous time random walk (CTRW). In the case of independently diffusing 
individuals, the dynamics appears to coincide with that of 
(dying and reproducing) Brownian walkers. In the strongly social case, the dynamics
coincides with that of non-migrating individuals. In both limits, the growth 
rate of the fluctuations becomes independent of the Hurst exponent of
the CTRW. The social behaviors that arise when transport in a population is
induced by a spatial distribution of random traps, have been analyzed.

\end{abstract}

\maketitle
\section{Introduction}
A population of Brownian walkers in the presence of processes of
birth and death, is known to undergo phenomena of spatial clustering \cite{zhang90,meyer96}.
It was argued that this effect could  
contribute in important way to the spatial inhomogeneity of the plankton distribution 
in the ocean \cite{young01}, and could have application in the modeling of algal blooms. 

The mechanism is of quite general nature, and can be 
seen as a simple ``bosonic'' realization of directed percolation \cite{paesens04}.
There are several situations, however, in which normal diffusion is not an appropriate
description of individual migration.  In the case of plankton, for example, the presence of
long lived vortices may lead to trapping phenomena and to the possibility of subdiffusive 
behaviors \cite{provenzale99}. Similar coupling of birth-death processes and subdiffusion
are expected to occur in the spreading of radioactive substances in geologic structures 
\cite{zoia08}.
On the other hand, the migration strategies of living organisms, over a large
range of scales, is often associated with long jumps \cite{leptos09,sims08},
and this may lead to superdiffusion \cite{klafter87}.  
Again, the generalization from Brownian to anomalously diffusing agents
has a close counterpart in the anomalous directed percolation studied in \cite{hinrichsen99},
in which jumps are modeled by L\'evy flights. 

It should be stressed that modelling transport in a random environment as a diffusive process,
even in the absence of demography, is a non trivial problem 
(see e.g. \cite{bouchaud90,hinrichsen00}).
The correlations induced by the spatial structure of the environment
are clearly lost if the individuals are treated as independent random walkers.
In the case of living organisms, these correlations could be seen as a form
of social behavior, and this leads naturally to the question of what is the impact 
of social behaviors, in the broad sense, on transport.

We have considered
a situation in which each individual, upon arrival at a certain location, fixes 
the time of next jump not only for itself, but also for all of its descendants,
and have shown that this leads to clustering phenomena 
analogous to those in a random environment. 
What we have found is that correlated behaviors analogous to those induced by the spatial
structure of the environment could be generated through inheritance, 
without the need to invoke any additional spatial interaction among individuals. 
This similarity between correlated behaviors is lost instead if
an ``antisocial'' condition is considered, in which each individual migrates independently,
following an internal clock. 

There is therefore a prescription problem in
the way newly generated individuals are made to jump, that is absent in the normal diffusion 
case. It is important to notice that this implies differences in the transport properties
of the population, that show up already at the level of one-individual statistics.
We mention that similar prescription problems were observed in \cite{yuste09}, which dealt with a
reaction -- anomalous diffusion system, with reaction of coagulation type: $A+A\to A$.

We shall consider a simple, spatially homogeneous population 
model, with Markovian births and deaths, and a migration dynamics 
of CTRW type \cite{montroll65,klafter87}.
Following \cite{young01}, we shall refer to the individuals in the simplified population model
as ``bugs''.
A single offspring will be assumed in each birth event,
with equal birth and death rates $\Gamma$ for stationarity.
Both superdiffusive and subdiffusive regimes will be considered, 
although, as it will soon appear, the degree of sociality of the bugs affects the long-time
population dynamics much more than the regime of anomalous diffusion.

\section{The model}
We briefly recall here the main properties of the CTRW model (more details in Appendix A).

The model is a generalization of the
random walk, in which the walker migrates through a sequence of independent jumps,
separated by randomly chosen waiting times.
The joint PDF (probability density function)
for the jump length $x$ and waiting time $t$ can be taken in the following form, 
in one dimension ($D=1$):
\beq
\psi(x,t)\sim\frac{\Delta t^\mu}{t^{1+\mu}\Delta x(t)}\exp\Big(-\frac{x^2}{2(\Delta x(t))^2}\Big),
\label{psi}
\eeq
where $0<\mu<1$ and $\Delta t\ll t$ fixes a microscale for the problem.
In the absence of birth and death processes, both subdiffusive and superdiffusive
behaviors may be generated, setting appropriately the jump length $\Delta x(t)$
\cite{klafter87}. Focusing on the case of a power-law dependence of the typical
jump length on the waiting time:
\beq
\Delta x(t)\simeq \Delta x(0)(t/\Delta t)^{\nu/2},
\qquad
t\gg\Delta t,
\label{Delta x}
\eeq
we have basically the following possibilities, provided we set $t\gg\Delta t$ and again neglect 
the effect of deaths \cite{klafter87,bouchaud90}:
\beq
&\nu\le\mu\Rightarrow \langle |x(t)-x(0)|^2\rangle\sim\kappa_\mu t^\mu,
\nonumber
\\
&\mu<\nu\Rightarrow \langle |x(t)-x(0)|^2\rangle\sim\kappa_\nu t^\nu,
\label{Hurst}
\eeq
where we have introduced the generalized diffusivity 
$\kappa_\alpha=[\Delta x(0)]^2/\Delta t^\alpha$.
We see that subdiffusion and superdiffusion are obtained respectively for $\nu<1$ and $\nu>1$.
However, as described in detail in Appendix A, the presence of demography prevents the long-time
asymptotics described in Eq. (\ref{Hurst}) to be ever reached. Thus, the long-time evolution of a 
population that was concentrated initially at a single spot, is described, in the presence of 
demography, by a normal diffusion process.

Of course, with or without demography, 
normal diffusion can be equally  generated by choosing a waiting time PDF that is not fat-tailed.


We shall study the CTRW model with births and deaths,
given an initial condition at time $t=0$, in which the bugs are uniformly
distributed in the domain: $\rho_1(x,0)=n_0$, where $\rho_1$ is the bug density.  As the
parameters in Eq. (\ref{psi}) are chosen independent of the coordinate, the density will
remain uniform also for $t>0$; in particular, imposing equality of the death and birth rates:
$\rho_1(x,t)=\rho_1(x,0)=n_0$. Demography however, through fluctuations, induces a
secular component in the higher order correlations, that is a manifestation of 
clustering.

To study the phenomenon, following  \cite{houch02}, we will consider the two-bug density
\beq
\rho_2(x_{1,2},t_{1,2})\equiv\rho_2(\x,\t)=n_0^2+\rho_{2c}(\x,\t),
\eeq
and we shall derive evolution equations for the connected
component $\rho_{2c}(\x,\t)$ in the two opposite regimes of antisocial and inheritance 
governed, strongly social bugs.  Notice that, contrary to \cite{houch02}, in which it was 
sufficient to consider the evolution of the one-time PDF $\rho_2(\x;t,t)$, the
non-Markovian nature of the CTRW will force us to study the full two-time 
statistical problem.


\section{The antisocial case}
In this case, the jump PDF is reinitialized at each birth event. 
The dynamics is that of
reactions in the presence of anomalous diffusion, that has been extensively studied
in the mean field regime (see e.g. \cite{schmidt07} and references therein).
Extension to two-bug statistics requires derivation of 
an evolution equation for the two-bug density $\rho_2(\x,\t)
\equiv\rho_2(x_{1,2},t_{1,2})$:
\beq
\rho_2(\x,\t)=\int_0^{t_1}\d\tau_1\int_0^{t_2}\d\tau_2\ 
\eta_2(\x,\t-\btau)\hat\Phi(\tau_1)\hat\Phi(\tau_2),
\label{master}
\eeq
where
$\eta_2(\x,\t-\btau)$ is essentially the rate of formation (birth plus arrival) of bug pairs in 
$(\x,\t-\btau)$ and the $\hat\Phi(\tau_{1,2})$ are the probabilities that the bugs arriving
or born in $(\x,\t-\btau)$ survive without jumping until $t_{1,2}$.
Since $\ex^{-\Gamma t}$ is the survival probability in a time $t$,
$\Phi(t)=\hat\Phi(t)\ex^{\Gamma t}$ is the no-jump probability;
from Eq. (\ref{psi}):
\beq
\Phi(t)=\int_t^\infty\d\tau\int\d x\ \psi(x,\tau)\sim (\Delta t/t)^\mu.
\label{Phi}
\eeq
Equation (\ref{master})
is the generalization of similar equations connecting the one-bug density $\rho_1$ and
arrival/birth rate $\eta_1$ derived in \cite{klafter87} and \cite{schmidt07}
(see Appendix A).

\begin{figure}
\begin{center}
\includegraphics[draft=false,width=7.5cm]{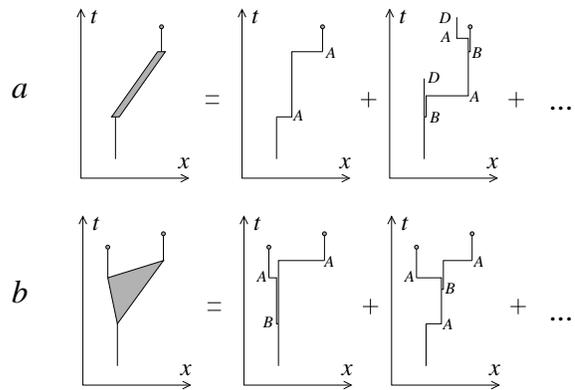}
\caption{
Contribution to the 1- and two-bug densities by different classes of family trees. 
Vertical lines ending with the small circle indicate a factor $\hat\Phi$. Letters
$A$, $B$, and $D$ indicate arrival, birth and death events. A line with a kink
ending in $A$ indicates the sequence of a waiting time and a jump, and is associated
with a factor $\hat\psi$, as indicated in Eq. (\ref{eta}). A birth event is associated
with a factor $\Gamma$.
}
\label{ctrwfig1}
\end{center}
\end{figure}

\begin{figure}
\begin{center}
\includegraphics[draft=false,width=8cm]{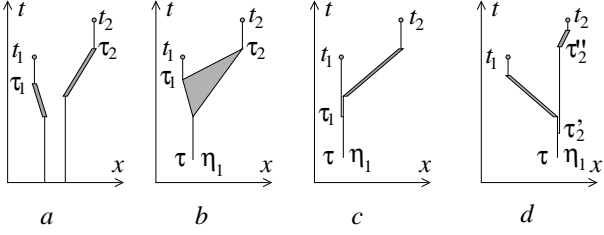}
\caption{
Family trees contributing to the two-bug density $\rho_2(\x,\t)$; 
($a$): disconnected part;
($b$): connected part $\rho_{2c}$; 
($c$) and ($d$) [in $(d)$ we have $t_2>\tau''_2>t_1$]: family trees acting as discrete source
for $\rho_{2c}$. Notice that chopping down branches in $(b)$, we end up with family trees 
in either form $(c)$ or $(d)$.
}
\label{ctrwfig2}
\end{center}
\end{figure}

The bug density $\rho_2(\x,\t)$ can be represented as a sum of contributions by family trees,
of the kind illustrated in Figs. \ref{ctrwfig1} and \ref{ctrwfig2}. The small circles
on top of the graphs identify the argument of the density function (the one-bug density
$\rho_1$ in the first line of Fig. \ref{ctrwfig1} and 
$\rho_2$ in the second) and the vertical lines ending in the
small circles identify the factors $\hat\Phi$. The part of the graphs that is obtained
chopping out these top lines accounts therefore for the rates $\eta_{1,2}$.

Clustering is accounted for by the connected part of $\rho_2$. The family trees 
responsible for this contribution are those summing to case $(b)$ in Fig. \ref{ctrwfig2}.
In this picture, graphs $(c)$ and $(d)$ in Fig. \ref{ctrwfig2} appear to play an important 
role. It turns out that they act as a discrete source in Eq. (\ref{master}), which we will
see is associated with the fact that any family tree in case $(b)$ of Fig. \ref{ctrwfig2} 
can be obtained adding or extending branches in a family tree in $(c)$ or $(d)$. 

The discrete nature
of graph $(c)$ can be understood, based on the direct parenthood relation implied
between bugs 1 and 2, which is an intrinsically discrete property
[in case $(d)$, this direct relation exists only when $t_1=t_2$].
These considerations prompt us 
to separate from the start discrete and continuous contributions
$\eta_2=\eta_2^{disc}+\eta_2^{cont}$, with $\eta_2^{disc}$ accounting
for graphs $(c)$ and $(d)$ in Fig. \ref{ctrwfig2}, and to rewrite the corresponding contribution
in Eq. (\ref{master}) as a source term:
\beq
s(\x,\t)=
\int_0^{t_1}\d\tau_1\int_0^{t_2}\d\tau_2\ 
\eta_2^{disc}(\x,\t-\btau)\hat\Phi(\tau_1)\hat\Phi(\tau_2).
\label{eq2}
\eeq
Let us evaluate explicitly this component. Let us start by
evaluating the increment $\Delta N^c_V$ 
that the family trees represented in case $c$ of Fig. \ref{ctrwfig2}, produce on
$N_V=\int_V\d x_2\rho_2(\x,\t)$, with $V$ some interval containing $x_1$.
The width $L_V$ of the interval $V$ will play the role of
a coarse graining length for $\rho_2$. 

To obtain $\Delta N^c_V$, we must calculate the product of the number of bugs generated
in $(x_1,\tau_1)$, surviving without jumps to $t_1$, and the number of bugs
originated after $\tau_1$ 
from the parent bug (PB) (plus the PB itself, if still alive) that are counted in 
$V$. The number of bugs generated by each PB in $(\tau_1,\tau_1+\d\tau_1)$,
that survive at $x_1$ without jumps till $t_1$, is 
$\Gamma\d\tau_1\hat\Phi(t_1-\tau_1)$.
Let us indicate with 
$G_V(t_2,\tau_1,\tau)$ the number of bugs originating from this same PB
after $\tau_1$ (again, including the PB itself, if still alive), that are
counted in $V$ at time $t_2$ ($\tau$ indicates the time of birth or arrival 
at $x_1$ of the PB). Taking into account the rate $\eta_1(x_1,\tau)$
of birth and arrival of bugs in $(x_1,\tau)$ and the probability $\hat\Phi(\tau_1-\tau)$
of their being available for parenthood in $(x_1,\tau_1)$, we obtain:
\beq
\Delta N^c_V&=&\Gamma\int_0^{t_1}\d\tau_1\int_0^{\tau_1}\d\tau\ \eta_1(x_1,\tau)
\hat\Phi(t_1-\tau_1)
\nonumber
\\
&\times&
\hat\Phi(\tau_1-\tau)G_V(t_2,\tau_1,\tau).
\label{eq3}
\eeq
Now, for $L_V$ large enough, all the descendants of the PB 
will remain in $V$, and,
provided the birth and death rates are equal: 
$G_V(t_2,\tau_1,\tau)\to 1$.
This causes
the two integrals in Eq. (\ref{eq3}) to decouple. We have 
$\int_0^{\tau_1}\d\tau\ \eta_1(x_1,\tau)\hat\Phi(\tau_1-\tau)=\rho_1(x_1,\tau_1)=n_0$,
the uniform one-bug density [this is just the evolution 
equation for the one-bug density; see Eq. (\ref{master1})]. 
From here we get the expression: 
\beq
\Delta N^c_V(\t)= n_0\Gamma\int_0^{t_1}\d\tau_1\hat\Phi(t_1-\tau_1).
\nonumber
\eeq
We proceed in identical way to calculate the contribution $\Delta N^d_V$ from 
graphs $(d)$ in Fig. \ref{ctrwfig2}. An analogous calculation leads to the result
\beq
\Delta N^d_V&=&\Gamma\int_0^{t_1}\d\tau'_2\int_0^{\tau_2'}\d\tau\ \eta_1(x_1,\tau)
\hat\Phi(t_1-\tau'_2)
\nonumber
\\
&\times&
\hat\Phi(\tau'_2-\tau)G_V(t_2,t_1,\tau'_2).
\nonumber
\\
&=& n_0\Gamma\int_0^{t_1}\d\tau'_2\hat\Phi(t_1-\tau'_2).
\nonumber
\eeq
From here we obtain for $\Delta N_V=\Delta N_V^c+\Delta N_V^d$, using Eq. (\ref{Phi}):
\beq
\Delta N_V(\t)= 2n_0\Gamma\int_0^{t_1}\d\tau_1\hat\Phi(t_1-\tau_1)\sim
n_0(\Gamma\Delta t)^\mu.
\label{eq4}
\eeq
We see that most bugs contributing in $(x_1,t_1)$ to $\Delta N_V(\t)$ are
born in an interval of width $\Delta t$ before $t_1$.
Thus, for $t_2-t_1\gg\Delta t$, the bugs in $V$ will have
dispersed (dying and reproducing) for a time $\simeq t_2-t_1$, and for Eq. (\ref{eq4}) to
hold, we need $L_V\gg X(t_2-t_1)$
where $X(t)$ is the typical separation at time $t$ of bugs originating from a
common ancestor at time zero.
We thus see that, if we are interested
in time scales $t_2-t_1\gg\Delta t$, and we are able to 
coarse grain $\rho_2$ at a scale $L_V\gg X(t_2-t_1)$, we can write for the
source in Eq. (\ref{eq2}):
\beq
s(\x,\t)\sim n_0(\Gamma\Delta t)^\mu\delta(x_1-x_2).
\label{source}
\eeq

Let us pass to analysis of $\eta_2^{cont}$, that can be decomposed 
into contributions from births ($B$) and arrivals ($A$):
\beq
\eta_2^{cont}=
\eta_2^{AA}+\eta_2^{AB}+\eta_2^{BA}+\eta_2^{BB},
\label{eta^cont}
\eeq
with $\eta_2^{AB}(x_1,t_1;x_2,t_2)= \eta_2^{BA}(x_2,t_2;x_1,t_1)$. The
components $\eta_2^{AA},\ldots$ obey equations that are generalizations of those for one-bug
\cite{klafter87,schmidt07} (see Appendix A).
It is convenient to isolate from the start the connected contributions to $\eta_2^{AA},\eta^{AB}_2$
and $\eta_2^{BA}$, meaning that the contribution from the initial condition 
$\rho_2(\x,0)$ is disregarded (contribution by graph $a$ in Fig.
\ref{ctrwfig2}).
This leads to the result:
\beq
\eta^{AA}_{2c}(\x,\t)&=&
\int_0^{t_1}\d\tau_1\int_0^{t_2}\d\tau_2\int\d^2y\ 
\eta_{2c}(\x-\y,\t-\btau)
\nonumber
\\
&\times&
\hat\psi(y_1,\tau_1)
\hat\psi(y_2,\tau_2),
\nonumber
\\
\eta^{AB}_{2c}(\x,\t)&=&n_0\Gamma
\int_0^t\d\tau\int\d y\ 
\eta_{1c}(x_1-y,t_1-\tau|x_2t_2)
\nonumber
\\
&\times&
\hat\psi(y,\tau),
\nonumber
\\
\eta^{BB}_{2c}(\x,\t)&=&\Gamma^2\rho_{2c}(\x,\t),
\label{eta}
\eeq
where $\hat\psi(x,t)=\psi(x,t)\ex^{-\Gamma t}$ is the PDF that the bug makes a jump of length
$x$ at time $t$ after birth or after a previous jump, and that therefore it did not
die in the meanwhile.
One more equation is needed for $\eta_{1c}(xt|x't')$, that is the rate of birth and arrival of 
bugs at $(x,t)$ given presence of a bug at $(x',t')$. 
This is the equation for the conditional density 
$\rho_{1c}(x_1t_1|x_2t_2)=n_0^{-1}\rho_{2c}(\x,\t)$:
\beq
\rho_{1c}(x_1t_1|x_2t_2)=\int_0^{t_1}\d\tau\ \eta_{1c}(x_1\tau|x_2t_2)
\hat\Phi(t_1-\tau).
\label{eq5}
\eeq
The integral equations (\ref{eta}) have a graphical interpretations as operations of
adding (through birth) or extending (through jumps) 
branches in the family trees of Figs. \ref{ctrwfig1}-\ref{ctrwfig2}. For
instance, the contribution $\eta_{2c}(\x-\y,\t-\btau)\hat\psi(y_1\tau_1)\hat\psi(y_2,\tau_2) 
\d^2\tau\d^2y$ to $\eta_{2c}(\x,\t)$ can be seen as the
operation of extending a branch from $(x_1-y_1,t_1-\tau_1)$ to $(x_1,t_1)$, and another
one from $(x_2-y_2,t_2-\tau_2)$ to $(x_2,t_2)$ in the family trees contributing
to $\eta_{2c}(\x-\y,\t-\btau)$. This substantiates the role of graphs $(c)$ and $(d)$ 
in Fig. \ref{ctrwfig2} as source for $\rho_2$ through Eqs. (\ref{master}) and (\ref{eta})
[notice that $\eta^{disc}_2$ still contributes in the RHS of the first and
second of Eq. (\ref{eta})].

\subsection{Dependence on the kind of diffusion}
The system of equations (\ref{master},\ref{eta^cont}-\ref{eq5}) can be solved by Fourier-Laplace
transform: $f(\x,\t)\to f_{\k\z}$.
Defining $\rho_{2c\k\z}=2\pi\delta(k_1+k_2)C_{k_1\z}$
and $s_{\k\z}=2\pi\delta(k_1+k_2)\sigma_{k_1\z}$, and indicating
$\hat z_{1,2}=\Gamma+z_{1,2}$, the result is, after little algebra (see Appendix B):
\beq
\{1-\psi_{k\hat z_1}\psi_{k\hat z_2}-\Gamma[\psi_{k\hat z_1}\Phi_{\hat z_2}+
\psi_{k\hat z_2}\Phi_{\hat z_1}]
\nonumber
\\
+\Gamma^2\Phi_{\hat z_1}\Phi_{\hat z_2}\}C_{k\z}=\sigma_{k\z}.
\label{eq6}
\eeq
We are going to consider separately the parameter ranges in Eq. (\ref{Hurst})
that would lead, in the absence of demography, to sub and super diffusion (the case of
normal diffusion is going to be discussed in Appendix C). We thus consider
$\mu<1$, $\nu=0$ for subdiffusion, and $\mu<1$, $\nu>1$ for superdiffusion.
Fourier-Laplace transforming Eq. (\ref{psi}) in the limit $k,z_{1,2}\to 0$, we obtain the
result, in the two cases:
\beq
\psi_{kz}=1-(\Delta\tilde t z)^\mu-(z\Delta\tilde t)^{\mu-\alpha}(\Delta\tilde x\,k)^2/2,
\label{psiz}
\eeq
where $\alpha=\mu,\nu$, respectively, in the subdiffusive and superdiffusive case,
and we have reabsorbed in the two parameters $\Delta\tilde t\sim\Delta t$, 
$\Delta\tilde x\sim\Delta x(0)$ any numerical coefficient arising in the Fourier-Laplace 
transform.
In the same way, we obtain from Eq. (\ref{Phi}) the expression for $\Phi_z$ valid in the
two cases:
\beq
\Phi_z=\Delta\tilde t^\mu z^{\mu-1}.
\label{Phiz}
\eeq
If we are interested in evaluating $\rho_{2c}(\x,\t)$ at separations $t_2-t_1\gg\Delta t$ and
$x_2-x_1\gg X(t_2-t_1)$, we can use Eq. (\ref{source}) to evaluate the source:
$\sigma_{k\z}\sim n_0(\Gamma\Delta t)^\mu/(z_1z_2)$; in the long-time regime 
$\Gamma t_{1,2}\gg 1$, we can Taylor expand in $z_{1,2}$, and Eq. (\ref{eq6}) becomes
in both superdiffusive and subdiffusive cases:
\beq
C_{k\z}\sim [(\bar z+\bar\kappa k^2)z_1z_2]^{-1} n_0\Gamma,
\label{eq7}
\eeq
where $\bar z=z_1+z_2$ and $\bar\kappa=\kappa_\alpha\Gamma^{1-\alpha}$,
(again, $\alpha=\mu,\nu$ in the subdiffusive and superdiffusive case).

The reason why both subdiffusive and superdiffusive regimes lead to the same
Eq. (\ref{eq7}) is that in the limit $\Gamma z_{1,2}\to 0$, we can expand
$\Phi_{\hat z_{1,2}}\simeq\Phi_\Gamma+\Phi'_\Gamma z_{1,2}$,
and similarly for $\psi_{k\hat z_{1,2}}$. In other words, the condition $\Gamma z_{1,2}\ll 1$
destroys any anomalous scaling with respect to $z_{1,2}$ in $\Phi_{\hat z_{1,2}}$
and $\psi_{k\hat z_{1,2}}$, and therefore also in $C_{k\z}$. This has the counterpart
in the original variables $(\x,\t)$ in the observation that for $\Gamma t_{1,2}\gg 1$,
the dominant factor in the individual bug dynamics is its own death, which prevents
the long waiting times responsible for anomalous scaling in Eq. (\ref{Hurst}) to be
ever reached (see Appendix A). 

Inverse Fourier-Laplace transform of Eq. (\ref{eq7}) leads to the final result
for $C(x,\t)\equiv\rho_{2c}(\x,\t)|_{x_2-x_1=x}$:
\beq
C(x,\t)\sim n_0X_\Gamma^{-1}(\Gamma\bar t)^{1/2}
f(x^2/(\bar\kappa\bar t)),
\label{C(x,t)}
\eeq
where $\bar t=(t_1+t_2)/2$, $X^2_\Gamma=\bar\kappa\Gamma^{-1}\sim (X(\Gamma^{-1}))^2$ 
is the square of the
typical bug displacement in a lifetime, and $f(x)$ is a function
equal to one at $x=0$ and decaying rapidly to zero at $|x|>1$. 
The scaling in $f$ shows that for $|t_2-t_1|\ll\bar t$ it is possible to set 
$L_V\gg X(|t_2-t_1|)$ and still have $L_V$ much smaller than the spatial 
scale of variation of $C$ at time $\bar t$. By continuity, this allows us to evaluate
$C(x,\t)$ at $x=0,t_1=t_2$, where the derivation leading to Eq. (\ref{source}) 
and then to (\ref{C(x,t)}) would not apply.

We notice the physical meaning of the parameter $\bar\kappa$, that is the only place 
in which memory of the initial scaling exponents of the process is preserved. This
is the effective diffusivity for the bugs in the presence of demography. Equation 
(\ref{C(x,t)}) is basically telling us that for
$\Gamma\bar t\gg$, the bugs behave as if they were simple Brownian walkers,
which execute at intervals $\Gamma^{-1}$
steps of order $X_\Gamma$.

\begin{figure}
\begin{center}
\includegraphics[draft=false,width=6cm]{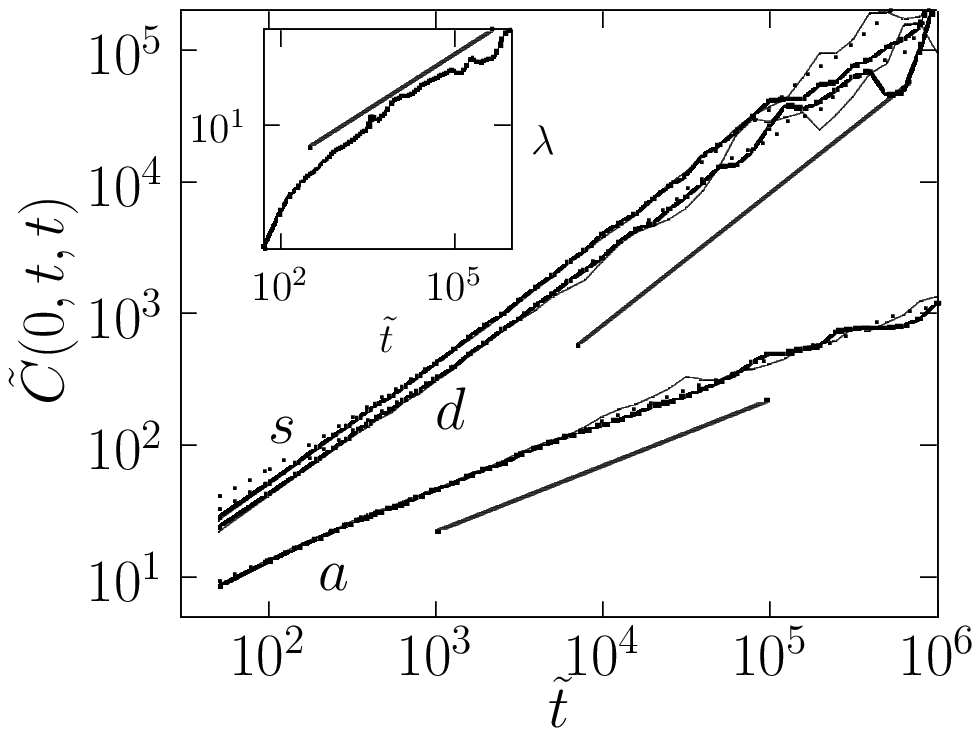}
\caption{
Result of numerical simulations in $D=1$ for $\mu=1/2$, $\nu=0$ (without demography, this
would correspond to a subdiffusive regime).
 Main graph: 
($a$) plot of 
$\tilde C=\Gamma^{-(1+\mu)/2}C$ vs. $\tilde t=t/\Delta t$ in the
antisocial case and fit by $\tilde t^{1/2}$;
($s$) plot of $\tilde C=\Gamma^{-1}C$ vs. $\tilde t=t/\Delta t$; in the social case;
($d$) plot of $\tilde C=\Gamma^{-1}C$ vs. $\tilde t=t/\Delta t$; in the case of
transport by a random field with correlation length $l_c=\Delta x/10$.
The fit of ($s$) and ($d$) is provided by $\tilde t$. In all three cases ($a$), ($s$) and ($d$),
dotted, heavy and thin lines correspond respectively to 
$\Gamma\Delta t=0.01,0.05,0.1$. 
Insert: plot of the correlation length $\lambda$ (units $\Delta x$) 
of $\tilde C$ in the antisocial case and fit by $\lambda\propto \tilde t^{1/2}$.  
In all cases: total population $N_0=2\times 10^5$; domain size $L=2\times 10^4\Delta x$.
In case $(d)$ we have put $\epsilon=10$ [see Eq. (\ref{occupation})].
}
\label{ctrwfig3}
\end{center}
\end{figure}

\begin{figure}
\begin{center}
\includegraphics[draft=false,width=6cm]{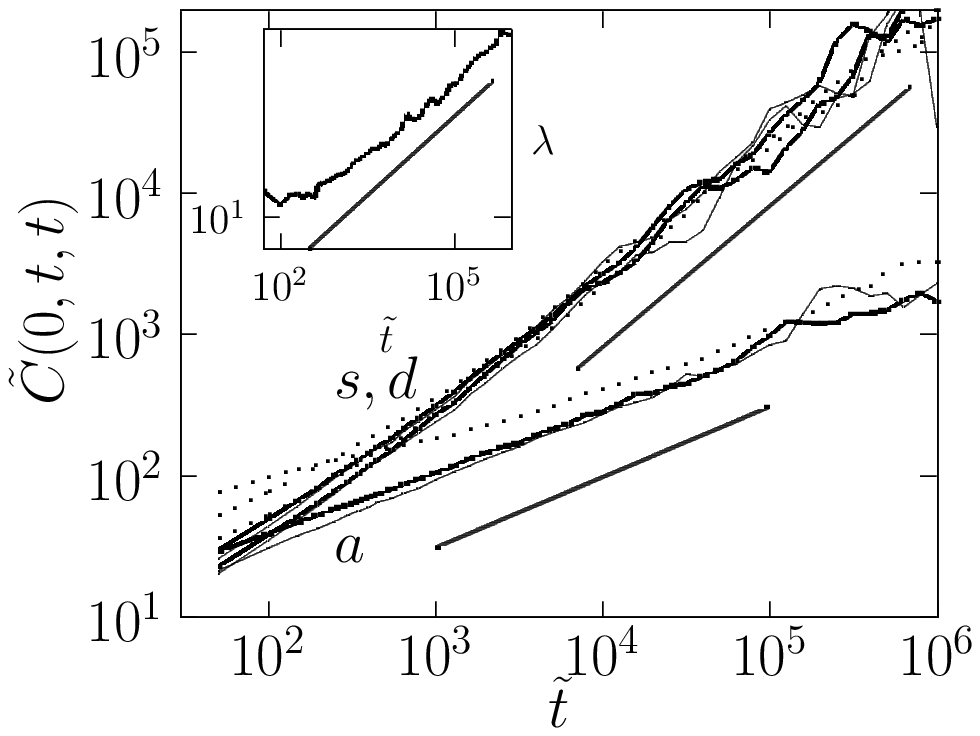}
\caption{
Same as Fig. \ref{ctrwfig3} in the case $\mu=1/2$, $\nu=3/2$ (without demography,
this would correspond to a superdiffusive regime).
}
\label{ctrwfig4}
\end{center}
\end{figure}

The scaling in Eq. (\ref{C(x,t)}) is the same as obtained in \cite{houch02},
in the case of individuals undergoing normal diffusion (Brownian bugs). 
As illustrated in Figs. \ref{ctrwfig3} and \ref{ctrwfig4}, this
result is confirmed by numerical simulations.
The similarity with the results in \cite{houch02} extends to $D>1$. 
Multiplying the right hand side (RHS) of Eq. (\ref{eq7}) by $k^{D-1}$ and then
inverse Fourier-Laplace transforming, we confirm in fact the result in \cite{houch02},
of logarithmic divergence of $C(x,\t)$ for $\bar t\to\infty$ for $D=2$ and no
divergence for $D=3$.


\section{The social case}
The analysis carried on so far illustrates the importance of initialization upon 
birth of the jumping PDF. One expects that anomalous behaviors may occur if a less
strong form of reinitialization for $\psi$ is adopted. However, it turns out that 
also imposing an excessive social constraint could result in anomalous scaling break-up.

We consider a situation in which the time of first jump for a bug 
is determined by the ancestor that 
initiated the colony to which the bug belongs. All the bugs in that colony
(and, if still alive, the ancestor itself) will jump therefore at the same time, 
independently.
It is interesting to notice that this social regime
is indistinguishable, at the one-bug level, from 
what would be obtained in the absence of demography.
The reason for this is that the number of individuals in a colony --
if the birth and death rates are equal -- will remain equal to one on the average.
The dispersal of the bugs, when their jump time arrives, will produce therefore
the same result, essentially, as what would be obtained from considering different realizations
of the initial bug in the absence of births and deaths.
(The same result would obviously be obtained considering the still tighter
constrain that all the bugs jump together as a single entity).

This situation leads to a strongly simplified population dynamics at the two-bug
level, as compared to the antisocial case. We can write in fact, setting
$t_2>t_1$:
\beq
\rho_2(\x,\t)&=&\int_0^{t_1}\d\tau_1\int_0^{t_2}\d\tau_2\ \eta_2^{AA}(\x,\t-\btau)
\nonumber
\\
&\times&\langle F(t_1-\tau)\rangle\langle F(t_2-\tau)\rangle
\nonumber
\\
&\times&\Phi(t_1-\tau_1)\Phi(t_2-\tau_2)+s(\x,\t),
\label{eq8}
\\
s(\x,\t)&=&\delta(x_1-x_2)\int_0^{t_1}\d\tau\ \eta_1^A(x_1,\tau)
\nonumber
\\
&\times&\langle F(t_1-\tau)F(t_2-\tau)\rangle \Phi(t_2-\tau),
\label{eq9}
\eeq
where $\eta_1^A$ is the rate of arrival of individual bugs, 
and $F(t)$ is the fluctuating number of descendants at time $t$ of a bug alive
at $t=0$. The variable $F$ describes a Galton-Watson process \cite{harris63}, 
which, for equal birth and death rates, is characterized by a constant average
$\langle F(t)\rangle=1$,
and a linearly growing correlation:
\beq
\langle F(t_1)F(t_2)\rangle=1+2\Gamma\min(t_1,t_2).
\label{Galton-Watson}
\eeq
Notice the absence in Eqs. (\ref{eq8}-\ref{eq9}) of hats on the $\Phi$'s, and of
contributions $AB,BA$ and $BB$ in $\eta_2$, as all demography is contained in the 
factors $F$. Absorption of demography in the factors $F$ and the fact that
$\langle F\rangle=1$ is what makes the dynamics identical at the one-bug
level, to the one of a simple CTRW in the absence of birth and death.
At the two-bug level of Eq. (\ref{eq8}), however, we still find a source $s(\x,\t)$,
that acounts for the fluctuations in the number of individuals in a given colony.
As illustrated in Eq. (\ref{eq9}), this quantity must be proportional to the arrival
rate of individual bugs at $x_1=x_2$, and to the probability that the colony does not
disperse before the final time $t_2$.

Let us analyze in detail the structure of this source term and start by evaluating
its long-time behavior. Notice that, thanks to the fact that $s(\x,\t)$ depends
only on space difference, we can write $s(\x,\t)=\sigma_k(\t)\delta(x_1-x_2)$.

Let us focus first on the equal time case and proceed by
Laplace transforming $\sigma_k(t,t)$.
We obtain:
\beq
{\cal L}_z[\sigma_k]=2\Gamma\eta^A_{1z}\Phi'_z\sim n_0\Gamma z^{-2}
\Rightarrow 
\sigma_k(t,t)\sim n_0\Gamma t,
\label{eq10}
\eeq
where use has been made of Eqs. (\ref{Phiz},\ref{eq9}) and (\ref{Galton-Watson}),
$\Phi'_z\equiv\d\Phi_z/\d z$, and of the result, generalizing Eq. (\ref{eq8}) to one-bug
statistics:
$n_0=\int_0^t\d\tau\eta_1^A(x,\tau)\Phi(t-\tau)\Rightarrow \Phi_z\eta_{1z}^A=n_0/z$
[compare with Eq. (\ref{master1})].
On the other hand we have, from Eqs. (\ref{eq9}) and (\ref{Galton-Watson}), for $t_2\gg t_1$:
$\sigma_k(\t)\sim\Gamma\Phi(t_2)\int_0^{t_1}\d\tau\eta_1^A(\tau)(t_1-\tau)$. Laplace transforming
in $t_1$, using again Eqs. (\ref{Phiz},\ref{eq9}-\ref{Galton-Watson})
and $\eta_{1z}^A=n_0/(z\Phi_z)$, we get:
\beq
{\cal L}_{z_1}[\sigma_k]\sim n_0\Gamma z_1^{-2-\mu}t_2^{-\mu}
\Rightarrow
\sigma_k(\t)\sim n_0\Gamma t(t_1/t_2)^\mu.
\label{eq11}
\eeq
We could interpolate Eqs. (\ref{eq10}-\ref{eq11}) to obtain, for $t_2\ge t_1$:
\beq
s(\x,\t)\sim n_0\Gamma t_1(t_1/t_2)^\mu\delta(x_1-x_2).
\label{source1}
\eeq
We notice at this point the important difference with respect to the antisocial regime.
While in that case fluctuation
build-up was produced as a balance between a constant $O((\Delta t)^\mu)$ forcing and
diffusion, in the present case, this build-up is realized directly in $s(\x,\t)$, 
which is here a macroscopic quantity \cite{note}. In the present case,
the only possible balance in Eq. (\ref{eq8}) is $\rho_2=s$ at $x_1=x_2$, which tells us 
that $\rho_2(\x,\t)$ must grow linearly in time for $x_1=x_2$. 

To be convinced of this result, let us write down explicitly the evolution equation for
$\rho_2$. We separate in the arrival rate $\eta_2^{AA}$ the
connected component $\eta_{2c}^{AA}$, which obeys an equation analogous to the first
of Eq. (\ref{eta}):
\beq
\eta^{AA}_{2c}(\x,\t)&=&
\int_0^{t_1}\d\tau_1\int_0^{t_2}\d\tau_2\int\d^2y\
\eta_{2c}(\x-\y,\t-\btau)
\nonumber
\\
&\times&
\psi(y_1,\tau_1)
\psi(y_2,\tau_2),
\label{eta^AA_2c}
\eeq
with $\eta_{2c}=\eta_{2c}^{AA}+\eta_2^{disc}$, and $\eta^{disc}_2$ providing the
discrete contribution to the social dynamics. This last contribution is associated with 
pairs of bugs that belong to a same colony and can be written in the form
\beq
\eta^{disc}_2(\x,\t)&=&2\Gamma\delta(t_1-t_2)\int_0^{t_1}\d\tau\int\d y\ \eta_1^A(y,\tau)
\nonumber
\\
&\times&
\psi(x_1-y|t_1-\tau)\psi(x_2-y|t_2-\tau),
\nonumber
\\
&\times&(t_1-\tau)\psi(t_1-\tau),
\label{eta^disc}
\eeq
where $\eta_1(y,\tau)$ gives the rate of arrival of the common ancestors at $(y,\tau)$,
and $2\Gamma(t_1-\tau)$, from Eq. (\ref{Galton-Watson}), 
is the average number of pairs of descendants per ancestor, 
available to jump at time $t_1=t_2$. [Of course, $\psi(t)$ is the waiting time PDF and 
$\psi(x|t)\equiv\psi(x,t)/\psi(t)$ is the jump length conditional PDF].

Substituting Eqs. (\ref{eta^AA_2c}) and (\ref{eta^disc}) into Eq. (\ref{eq8}),
exchanging the order of the convolutions and setting again $\langle F\rangle=1$,
we obtain the expression:
\beq
\rho_{2c}(\x,\t)&=&\int_0^{t_1}\d\tau_1\int_0^{t_2}\d\tau_2\int\d^2y\ \psi(y_1,\tau_1)
\psi(y_2,\tau_2)
\nonumber
\\
&\times&K(\x-\y,\t-\btau)+s(\x,\t),
\nonumber
\eeq
where $K=\rho_{2c}+\mathfrak{s}-s$, and
\beq
\mathfrak{s}(\x,\t)&=&\int_0^{t_1}\d\tau_1\int_0^{t_2}\d\tau_2\ 
\eta_2^{disc}(\x,\t-\btau)
\nonumber
\\
&\times&
\Phi(\tau_1)\Phi(\tau_1).
\label{frak s}
\eeq
From here we reach the final result:
\beq
\rho_{2c}(\x,\t)=s(\x,\t)+\bar\rho_{2c}(\x,\t),
\label{rho_2c}
\eeq
where
$\bar\rho_{2c}$ obeys the equation
\beq
\bar\rho_{2c}(\x,\t)&=&\int_0^{t_1}\d\tau_1\int_0^{t_2}\d\tau_2\int\d^2y\ \psi(y_1,\tau_1)
\psi(y_2,\tau_2)
\nonumber
\\
&\times&\bar\rho_{2c}(\x-\y,\t-\btau)+\bar s(\x,\t),
\label{bar rho}
\eeq
and
\beq
\bar s(\x,\t)&=&\int_0^{t_1}\d\tau_1\int_0^{t_2}\d\tau_2\int\d^2y\ \psi(y_1,\tau_1)
\psi(y_2,\tau_2)
\nonumber
\\
&\times&\mathfrak{s}(\x-\y,\t-\btau)
\label{bar s}
\eeq
is a source term that, as can be checked from 
Eqs. (\ref{eta^disc},\ref{frak s}) and (\ref{bar s}),
is continuous with respect to $x_1-x_2$.

Comparing with Eqs. (\ref{source1}) and (\ref{bar rho}), 
we thus see that the two terms on the RHS of Eq. (\ref{rho_2c}) account respectively 
for the singular and regular components of $\rho_{2c}$. The singular component $s(\x,\t)$, 
in particular, coincides with what would be observed 
in a population that does not migrate (in fact, its dynamics is that of a Galton-Watson
process). The only place in the singular component, where migration seems to have 
some effect, is the structure of time correlations, as accounted for by
the power-law decay in Eq. (\ref{source1}) [compare with Eq. (\ref{Galton-Watson})].
As illustrated in Figs. \ref{ctrwfig3} and \ref{ctrwfig4}, the clustering dynamics
at long times is dominated by the singular component $s(\x,\t)$ in both the superdiffusive
and the subdiffusive case. 

We stress that these results are conditioned to the waiting
time PDF $\psi(t)$ being fat-tailed. If this condition were not satisfied, it is possible
to see that $\rho_{2c}$ would not have a singular component and the dynamics
would become that of Brownian bugs \cite{note}.

\section{Population dynamics in a random environment}
The kind of social behavior discussed in the previous section is very close to what
one could expect if the bug migration were determined by a spatial arrangement 
of random traps. The opening time of the trap determines the time of dispersal
of the bugs that are located in it. When a trap opens, all the bugs
jump independently and remain caught either in the same trap (whose opening
time is now updated) or in some other trap elsewhere. 

The important difference, with respect to the case of inheritance produced social behaviors,
is that now, colonies that were initiated
by bugs that happened to be caught in the same trap,
will share an identical dispersal time.
In the case considered in the previous section, the dispersal time of colonies sharing the
same location in space was in general different.

To verify the possibility of discrepancies between the two mechanisms for
social behavior, a series of numerical simulation have been carried on in 
$D=1$ and $D=2$, with  random traps whose opening times are
independently distributed with PDF $\psi(t)\propto t^{-1+\mu}$. The jump 
length PDF for the bugs $\psi(x|t)=\psi(x,t)/\psi(t)$ is the one provided 
by Eq. (\ref{psi}).

\begin{figure}
\begin{center}
\includegraphics[draft=false,width=7.5cm]{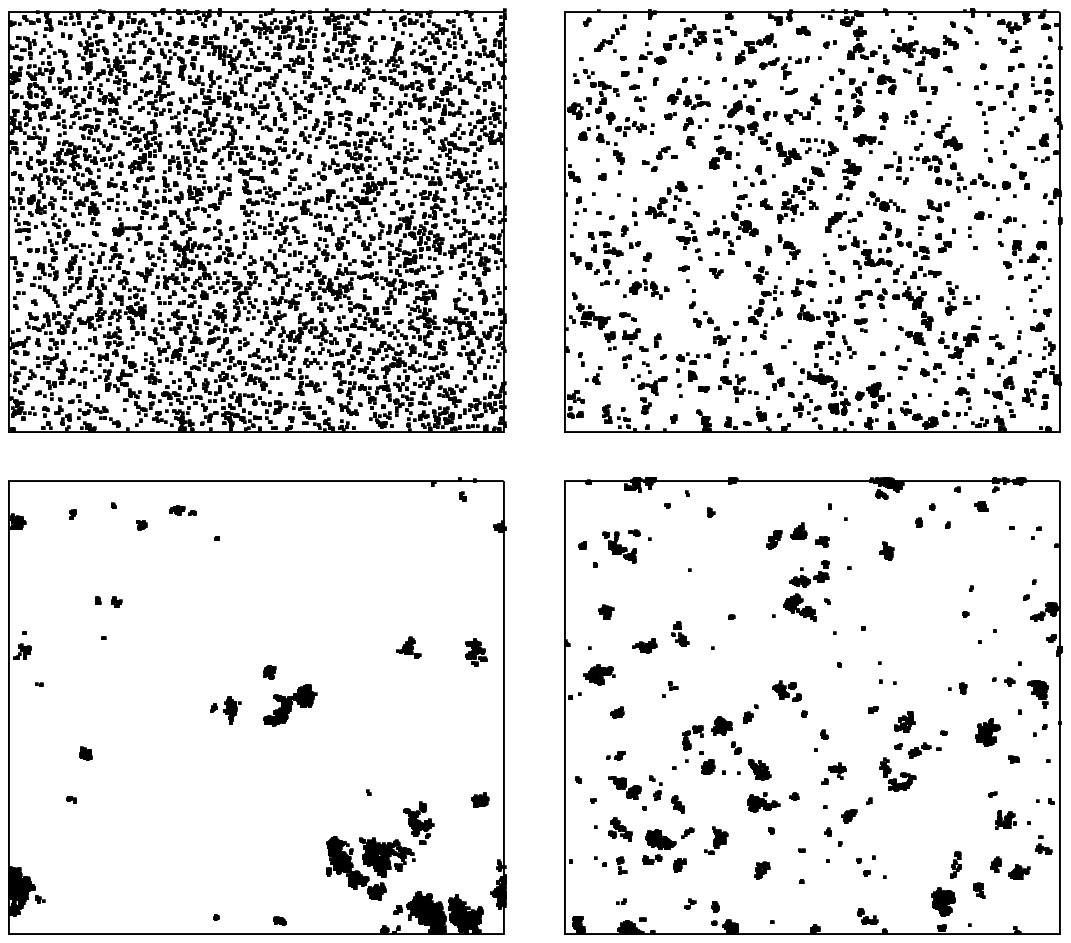}
\caption{
Sequence of snapshots of the population  
spatial distribution from a numerical simulation 
of bugs in a $D=2$ random environment. Going clockwise from top left:
$t/\Delta t=4\times 10^3,4\times 10^4,4\times 10^5,4\times 10^6$. 
Parameters of the simulation: $\mu=1/2$, $\nu=3/2$ (would be superdiffusive case);
$\Gamma\Delta t=0.01$; $\kappa_0=10^{-3}(\Delta x(0))^2/\Delta t$.
Domain size in units $\Delta x(0)$:
$1400\times 1400$. Correlation length for random field: $l_c=14\Delta x(0)$.
Total population: $N_0=2\times 10^5$; occupation number: $\epsilon=10$.
}
\label{ctrwfig5}
\end{center}
\end{figure}

In all cases considered, periodic boundary conditions were utilized, and the total number
of bugs was kept fixed, varying from the outside the birth and death rates in the population. 
(The meaning of a constant population constraint is to subtract any global finite population effect
from the local fluctuations associated with clustering \cite{note1}). 
In addition to 
this, in $D=2$, the possibility of a small normal diffusivity $\kappa_0$ 
acting beside the jumps,
has been considered, to mimick the effect of individual small scale motion 
of the bugs. 
(In the case of plankton, this additional component may be interpreted as due
to individual swimming, while the long jumps are associated with eddy break-up and
and transport of plankton patches away, along strain filaments).
A picture of what would occur in this case is presented in 
Fig. \ref{ctrwfig5}.

It is clear that 
the discrepancy between different social mechanisms
will be negligible if the population is so dilute that on the average, at most
one colony (and therefore one bug) is present at a given time in a given trap.
Given a typical size $l_c$ of the trap, an important parameter is therefore
the product 
\beq
\epsilon=n_0l^D_c
\label{occupation}
\eeq
that gives the typical occupation number of a trap. 

\begin{figure}
\begin{center}
\includegraphics[draft=false,width=6cm]{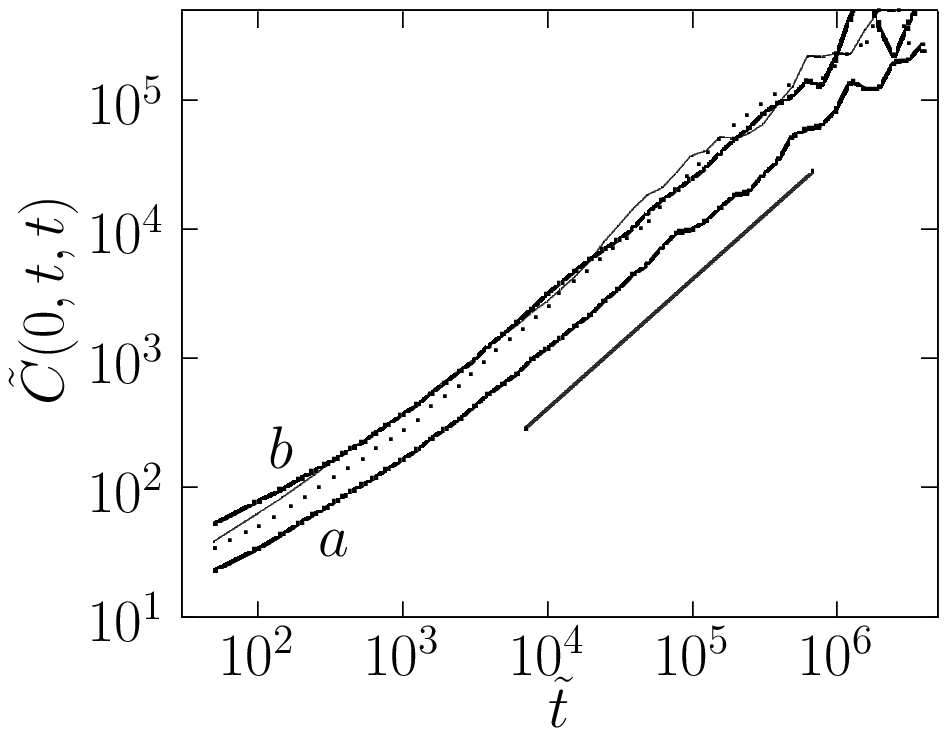}
\caption{
Plot of  $\tilde C=\Gamma^{-1}C$ vs. $\tilde t=t/\Delta t$ and fit by $\tilde t$, in the
case of transport by a random field. $(a)$: $D=2$, the same case as Fig. \ref{ctrwfig5}; 
the fluctuation amplitude $C(0;t,t)$ was coarse grained at scale $l_c$. The case 
$\kappa_0=0$ leads to a profile that is indistinguishable from the one in figure.
$(b)$ $D=1$; again a would be superdiffusive case, with $\mu=1/2$, $\nu=3/2$ and
$\Gamma\Delta t=0.01$;
the three curves correspond to to $\epsilon=1$ 
(heavy line), $\epsilon=20$ (thin line), $\epsilon=200$ (dotted line); for the rest,
the same parameter as case $(d)$ in Figs. \ref{ctrwfig3} and \ref{ctrwfig4} were utilized.
A slope $\propto\tilde t$ is shown for comparison.
}
\label{ctrwfig6}
\end{center}
\end{figure}

What we have found, in both $D=1$ and $D=2$, 
is that clustering is remarkably insensitive
to whether the population is concentrated or diluted, or to the presence of small scale
motions that cause bug leakages from the traps. As illustrated
in Fig. \ref{ctrwfig6}, a change of more than two orders of magnitude in $\epsilon$ does not
produce any appreciable change in the fluctuation amplitude $C(0;t,t)$. 
In the same way, as can be seen in Figs. \ref{ctrwfig3} and \ref{ctrwfig4},
the difference in the profile for $C(0;t,t)$ in the two cases in which sociality
is produced by inheritance and spatially extended random traps, are negligible.
The same result holds both in the subdiffusive and the superdiffusive case.

An explanation for the insensitivity of clustering to the value of the parameter
$\epsilon$, and to the mechanism for sociality, is the fact the clustering behavior
accounted for by Eqs. (\ref{source1}) and (\ref{rho_2c}) is already extremal. 
The maximal fluctuation growth, corresponding to individuals
that do not migrate, as described by the Galton-Watson regime of Eq. 
(\ref{Galton-Watson}), is already achieved in the inheritance governed
dynamics of the previous section. Passing to a random trap governed dynamics,
further diminishes the degree of dispersal of the bugs and maintains the
clustering on the maximal level $C(0;t,t)\propto t$. Increasing the
value of the parameter $\epsilon$ goes in the same direction and does
not modify the result.

Geometrically, as accounted for by Eqs. (\ref{source1}) and (\ref{rho_2c}), 
this corresponds to a situation in which, unless some
small scale migration is assumed, the typical cluster size is zero. 
Repetition of the $D=2$ simulation of Fig. \ref{ctrwfig5} with $\kappa_0=0$
can be shown to lead, in fact, to a bug distribution in the long time limit, that
is a discrete set of isolated towers, as it would occur with bugs that do not migrate
at all \cite{young01}. The same can be seen to occur also in the $D=1$ simulations,
in both the inheritance and random trap governed cases.
This contrasts with the result in the antisocial case (see insert in Figs. \ref{ctrwfig3}
and \ref{ctrwfig4}), which describes cluster whose size $\lambda$ is a growing function 
of time.

\section{Conclusions}
The main result of the present analysis is that social behaviors can play a
strong role in the transport properties of a population, whose individuals
migrate following a CTRW strategy.
An interesting aspect, with potentially useful
application, is that 
enforcing a simple inheritance-social constraint on the bugs, 
is sufficient 
to replicate the dynamics of
a population in a random environment
(say, an assembly of random traps).
This means that it is not necessary to impose spatial
interactions among the CTRW-bugs to mimick the common response 
of nearby individuals to the environment. Nevertheless,
a social constraint is necessary to guarantee that the dynamics
of CTRW-bugs and that of individuals in a field of random traps be identical.
Antisocial bugs, for instance, do not satisfy this property and their dynamics
coincides instead with that of Brownian bugs.
In particular, as in the Brownian case of \cite{houch02}, 
antisocial bugs are characterized by reduced clustering in 
$D>1$, with fluctuations that grow only logarithmically ($D=2$) or
saturate to a constant level ($D=3$).


The clustering behavior in the social case  (and therefore also in the case of 
a random environment) coincides with what would be
observed in the case of bugs that do not migrate. 
The growth of the fluctuation amplitude
is described by a Galton-Watson process independently of the dimensionality of
the problem, and the size of the clusters is determined by eventual 
small scale motions of the individuals. (This is opposite to the behavior
of antisocial and Brownian bugs.
where the clusters are produced by a competition between migration and
demography). In consequence of this, no clustering reduction with
increasing dimensionality occurs in the social case.

Thus, in realistic situations of a population dispersed in a bi- or three-dimensional 
random environement, a transport dynamics governed by random traps with long permanence
times, will lead to much stronger clustering behaviors than it would occur if the
transport dynamics were of the  normal diffusion type. This result is independent of the
kind of transport, either superdiffusive or subdiffusive, that would result if no
birth and death effects were present. The only important constraint is 
that the waiting time PDF in the CTRW (or in the random traps) be fat-tailed. 
On the contrary, the strong role of sociality is
lost in the case of Brownian bugs. Also, transport by an assembly of random traps
with exponentially distributed opening times, turns out to be described by a Brownian bug
dynamics, without the need of imposing social constraints of any sort.

In all cases, demography appears to be very good at destroying anomalous scaling
behaviors. Antisocial behaviors bring the dynamics back
to that of Brownian bugs; strongly social behaviors
bring it back to that of non-migrating individuals. 
In the antisocial case, 
anomalous scaling could possibly be recovered for $\bar t\ll\Gamma^{-1}$.
From Eq. (\ref{C(x,t)}),
we see however that fluctuations become relevant, i.e. $C(x,t,t)\sim n_0^2$, only
when $\Gamma\bar t\sim (n_0X_\Gamma)^2$,
which is small unless the density is so low that the typical bug separation is at least 
of the order of their displacement in a lifetime.

The question remains open on whether anomalous
scaling could be obtained from regimes half-way between the strongly social and
antisocial cases that have been considered so far.
Anomalous scaling could arise as well from social behaviors that do not
originate from inheritance, but are the result 
of more general spatial interactions between the bugs, as discussed e.g.
in \cite{heinsalu10}. Similarly, the question remains open on the effect of
demography on bugs in which anomalously diffusive behaviors are produced 
by means different from a CTRW, say, a fractional Brownian motion \cite{beran}
or a sequence of L\'evy flights \cite{heinsalu10}.




\acknowledgements I wish to thank Silvano Ferrari for his contribution to the initial 
part of this research. This research was funded in part by Regione Autonoma della Sardegna.

\appendix
\section{Anomalous diffusion and demography in the antisocial case}
To check for the presence of anomalous diffusion behaviors, we study the scaling of
the bug density $\rho_1(x,t)$, given the initial condition $\rho_1(x,0)=\delta(x)$. This
choice of initial conditions
makes $\rho_1(x,t)$ coincide with the PDF of finding 
the original bug or one of its descendants at position $x$ at time $t$.

The integral equation satisfied by $\rho_1$ has a form analogous to Eq. (\ref{master})
\cite{klafter87,schmidt07}:
\beq
\rho_1(x,t)=\int_0^t\d\tau\ \eta_1(x,\tau)\hat\Phi(t-\tau),
\label{master1}
\eeq
where $\eta_1$ is the rate of arrival and birth of individual bugs at $(x,t)$, and 
$\hat\Phi(t)=\ex^{-\Gamma t}\Phi(t)$ is the probability that the bug does not
jump or die in a time $t$ [see Eq. (\ref{Phi})]. Let us decompose as in Eq. (\ref{eta^cont}),
$\eta_1$ into contributions from arrival and birth:
$\eta_1=\eta_1^A+\eta_1^B$. The arrival contribution $\eta_1^A(x,t)$ is obained summing
over all the bugs that where born or arrived at a previous time $\tau$ at any position $y$ and
from there jumped at $x$ at time $t$. Following \cite{klafter87}, it is possible to include in 
$\eta_1^A(x,t)$ an initial condition $\rho(x,0)$, so that we can write, for $t\ge 0$:
\beq
\eta_1^A(x,t)&=&\int_0^t\d\tau\int\d y\ \eta_1(y,\tau)\hat\psi(x-y,t-\tau)
\nonumber
\\
&+&\rho(x,0)\delta(t),
\label{arrival}
\eeq
while $\eta(x,t)=0$ for $t<0$. The factor $\hat\psi(x,t)=\ex^{-\Gamma t}\psi(x,t)$ is the
PDF that the bug jumps at at a time $t$ 
after a previous arrival or birth, that the length of the
jump is $x$, and that the bug is still alive at the time of the jump. 
The birth rate $\eta_1^B(x,t)$ satisfies instead:
\beq
\eta_1^B(x,t)=\Gamma\rho_1(x,t).
\label{birth}
\eeq
Substituting Eqs. (\ref{birth}) and (\ref{arrival}) into Eq. (\ref{master1}), we obtain,
after inverting the order of the convolutions, the evolution equation:
\beq
\rho_1(x,t)&=&\int_0^t\d\tau\Big[\int\d y\ \rho_1(y,\tau)\hat\psi(x-y,t-\tau)
\nonumber
\\
&+&\Gamma\rho_1(x,\tau)
\hat\Phi(t-\tau)\Big]+\rho(x,0)\hat\Phi(t).
\label{rho_1}
\eeq
Fourier transforming in space and Laplace trasforming in time, Eq. (\ref{rho_1}) becomes,
imposing the initial condition 
$\rho_1(x,0)=\delta(x)$:
\beq
\rho_{1kz}=\frac{\Phi_{\hat z}}{1-\psi_{k\hat z}-\Gamma\Phi_{\hat z}},
\nonumber
\eeq
where, as in Eq. (\ref{eq6}), $\hat z=\Gamma+z$. From here we obtain, in the
the regime $\hat z\Delta t\ll 1$, using Eqs. (\ref{psiz}) and (\ref{Phiz}):
\beq
\rho_{1kz}\simeq\frac{1}{z}-\frac{1}{2}\kappa_\alpha\hat z^{1-\alpha}z^{-2}k^2
\eeq
where $\kappa_\alpha\sim\Delta\tilde x^2\Delta\tilde t^{-\alpha}$, with $\alpha=\mu,\nu$, 
respectively, in the subdiffusive and superdiffusive case. 
The variance 
$X^2(t)$ of the PDF $\rho_1$ is obtained as the inverse Laplace transform of the quantity 
\beq
-\frac{\partial^2\rho_{1kz}}{\partial k^2}\Big|_{k=0}=
\begin{cases}
\kappa_\alpha z^{-1-\alpha}, &\Gamma\ll z\ll\Delta\tilde t^{-1}
\\
\bar\kappa z^{-1},&z\ll\Gamma
\end{cases}
\eeq
where $\bar\kappa=\kappa_\alpha\Gamma^{1-\alpha}$. We obtain the final result
\beq
X^2(t)\sim
\begin{cases}
\kappa_\alpha t^\alpha,& \Delta\tilde t\ll t\ll\Gamma^{-1}
\\
\bar\kappa t,& t\gg\Gamma^{-1}
\end{cases}
\eeq
and we see that demography kills anomalous scaling for $\Gamma t\gg 1$.

\section{Equations for the two-bug statistics in the antisocial case}
Let us fill the steps to go from Eqs. (\ref{master},\ref{eta^cont}-\ref{eq5}) to Eq. (\ref{eq6}).
Substituting Eqs. (\ref{eta}) into the connected part of Eq. (\ref{master}), we find:
\beq
\rho_{2c}(\x,\t)&=&\int_0^{t_1}\d\tau_1\int_0^{t_2}\d\tau_2\int\d^2 y\
\rho_{2c}(\y,\btau)
\nonumber
\\
&\times&
\hat\psi(x_1-y_1,t_1-\tau_1)
\hat\psi(x_2-y_2,t_2-\tau_2)
\nonumber
\\
&+&\int_0^{t_1}\d\tau_1\int_0^{t_2}\d\tau_2\Big[[\eta_{2c}^{AB}(\x,\btau)
+\eta_{2c}^{BA}(\x,\btau)
\nonumber
\\
&+&\eta_{2c}^{BB}(\x,\btau)\Big]
\hat\Phi(t_1-\tau_1)\hat\Phi(t_2-\tau_2)
\nonumber
\\
&+&s(\x,\t),
\label{eq18}
\eeq
where use has been made of Eq. (\ref{eq2})  to eliminate $\eta^{disc}$ from the
expression.
On the other hand, we can write, using the second of Eq. (\ref{eta}) and Eq. (\ref{eq5}):
\beq
&&\int_0^{t_1}\d\tau_1\int_0^{t_2}\d\tau_2\ \eta_{2c}^{AB}(\x,\btau)
\hat\Phi(t_1-\tau_1)\hat\Phi(t_2-\tau_2)
\nonumber
\\
&&=n_0\Gamma\int_0^{t_2}\d\tau_2
\int_0^{t_1}\d\tau_1\int_0^{\tau_1}\d\tau'_1\int\d y_1\
\hat\Phi(t_2-\tau_2)
\nonumber
\\
&&\times
\eta_{1c}(y_1\tau_1|x_2\tau_2)
\hat\psi(x_1-y_1,\tau_1-\tau'_1)\hat\Phi(t_1-\tau_1)
\nonumber
\\
&&=\Gamma\int_0^{t_2}\d\tau_2
\int_0^{t_1}\d\tau_1\int\d y_1
\ \hat\Phi(t_2-\tau_2)
\nonumber
\\
&&\times
\rho_2(y_1\tau_1;x_2\tau_2) \psi(x_1-y_1,t_1-\tau_1).
\label{eq18.5}
\eeq
Substituting Eq. (\ref{eq18.5}) with the third of Eq. (\ref{eta}) into Eq. (\ref{eq18}),
we get the result, after inverting the order of the convolutions:
\beq
\rho_{2c}(\x,\t)&=&\int_0^{t_1}\d\tau_1\int_0^{t_2}\d\tau_2\Big\{\int\d^2 y\
\rho_{2c}(\y,\btau)
\nonumber
\\
&\times&\hat\psi(x_1-y_1,t_1-\tau_1)\hat\psi(x_2-y_2,t_2-\tau_2)
\nonumber
\\
&+&\Gamma\int\d y_1\ \rho_{2c}(y_1,x_2;\btau)\hat\psi(x_1-y_1,t_1-\tau_1)
\nonumber
\\
&\times&\hat\Phi(t_2-\tau_2)
+\Gamma\int\d y_2\ \rho_2^c(x_1,y_2;\btau)
\nonumber
\\
&\times&\hat\psi(x_2-y_2,t_2-\tau_2)\hat\Phi(t_1-\tau_1)
\nonumber
\\
&+&\Gamma^2\rho_2^c(\x,\btau)\hat\Phi(t_1-\tau_1)\hat\Phi(t_2-\tau_2)\Big\}
\nonumber
\\
&+&s(\x,\t),
\label{eq19}
\eeq
which, Fourier transforming in space and Laplace transforming in time, 
gives  Eq. (\ref{eq6}).

\section{Application to the case of Brownian bugs}
Using Eqs. (\ref{eq6}) and (\ref{source}) in the
Brownian bug case is like using a shotgun to take down a mosquito. Nevertheless,
it allows one to test the techniques derived 
in the present paper on the known results in \cite{houch02}.

In the case of Brownian bugs we have,
in place of Eqs. (\ref{psi}) and (\ref{Phi}):
\beq
\psi(x,t)=
\frac{2^{1/2}}{\pi^{1/2}\Delta t\Delta x}
\exp\Big(-\frac{t}{\Delta t}-\frac{x^2}{2\Delta x^2}\Big)
\eeq
and
\beq
\Phi(t)=\exp(-t/\Delta t).
\label{C2}
\eeq
From here we can write, in place of Eqs. (\ref{psiz}) and (\ref{Phiz}):
\beq
\psi_{kz}=\frac{\exp(-(\Delta x k)^2/2)}{1+\Delta tz};
\qquad
\Phi_z=\frac{\Delta t}{1+\Delta tz}.
\label{C3}
\eeq
Substituting Eq. (\ref{C2}) into Eq. (\ref{eq4}), we obtain, in place of
Eq. (\ref{source}):
\beq
s(\x,\t)=2n_0\Gamma\Delta t\delta(x_1-x_2).
\label{C4}
\eeq
Substituting Eqs. (\ref{C3}) and (\ref{C4}) into Eq. (\ref{eq6}) and Taylor expanding
around $k,\z=0$, we find in place of Eq. (\ref{eq7}):
\beq
C_{k\z}=2[(\bar z+\kappa k^2)z_1z_2]^{-1} n_0\Gamma,
\label{C5}
\eeq
where $\bar z=z_1+z_2$, and
$\kappa=\Delta x^2/\Delta t$ is the diffusion of the Brownian walkers.
Contrary to the case of Eq. (\ref{eq7}), the transport part of
the dynamics, accounted for by the diffusivity $\kappa$, is independent of 
demography. We see that Eq. (\ref{C5}) is the Laplace transform of the
forced heat diffusion equation 
\beq
\partial_tC(x,t)=\kappa\partial_x^2C(x,t)+2n_0\Gamma\delta(x),
\label{C6}
\eeq
where $C(x,t)\equiv C(x;t,t)$.

It is possible to see that Eq. (\ref{C6}) corresponds to the continuous limit
of the result in \cite{houch02}. Rather than taking the continuous limit
of that result, we take the limit directly on the original problem on 
the lattice. 

Let the discretization in time be given by $\Delta t$ and
the one in space by $\ell>\Delta x$. Thus $C(x,t)=\ell^{-2}\langle N(0,t)N(x,t)\rangle$,
with $N(x,t)$ the instantaneous number of bugs in the cell $x$.
We can easily fix the jump probability, to obtain a diffusive dynamics with 
diffusivity $\kappa=\Delta x^2/\Delta t$. Analysis of the source term in Eq. (\ref{eq6})
requires a little more work.  Let us indicate by $\Delta C(x,t)$ the
contribution to $C(x,t+\Delta t)-C(x,t)$ from demography. Again, we easily see
that for $x\ne 0$, due to independence of the birth and death processes,
we have 
$\ell^2\Delta C(x,t)=\langle N(0,t)\Delta N(x,t)\rangle+\langle\Delta N(0,t)N(x,t)\rangle=0$.
If $x=0$, however, we have that the quantity $\langle\Delta N^2(0,t)|N(0,t)\rangle$
is non-zero. Introducing the indicator function $\delta_i=0,1,2$ depending on whether
bug $i$ dies, remains inactive, or gives birth, we have in fact a contribution
to $\langle\Delta N^2(0,t)|N(0,t)\rangle$:
\beq
\sum_{ij}^{N(0,t)}\langle\delta_i\delta_j\rangle= 2N(0,t)\Gamma\Delta t,
\nonumber
\eeq
where we have used the fact that $\langle\delta_i\delta_j\rangle=0$ for $j\ne j$, and
$\delta_i=0,1,2$ with probabilities $\Gamma\Delta t$, $1-2\Gamma\Delta t$ and $\Gamma\Delta t$.
Exploiting the fact that $\langle N(0,t)\rangle=n_0\ell$, we find in the end
\beq
\Delta C(0,t)=2n_0\Gamma\Delta t\ell^{-1},
\nonumber
\eeq
which leads in the continuous limit to the expression for the source provided in Eq. (\ref{C6}).


\begin{thebibliography}{10}
\bibitem{note} This property is lost in the case of Brownian bugs, in which 
case Eq. (\ref{eq10}) becomes $\sigma_k(t,t)\sim n_0\Gamma\Delta t$.
\bibitem{note1} For a total population $\langle N(t)\rangle=N_0$ the global fluctuation
amplitude can be estimated using Eq. (\ref{Galton-Watson}): $\langle [N(t)]^2\rangle=
N_0^2+2N_0\Gamma t$.
\end{thebibliography}
\end{document}